\documentstyle[prl,aps,epsf,twocolumn]{revtex}
\def\break#1{\pagebreak \vspace*{#1}}
\begin{document}
\def\bea{\begin{eqnarray}}
\def\eea{\end{eqnarray}}
\def\a{\alpha}
\def\D{\langle l \rangle}
\def\p{\partial} 
\draft
\title{Residence Time Distribution for a Class of Gaussian Markov
Processes}
\author{Abhishek Dhar $^1$ and Satya N. Majumdar $^2$}
\address{$^1$ Physics Department, Indian Institute of Science,
Bangalore 560012, India. \\ $^2$ Theoretical Physics Group, Tata
Institute of Fundamental Research, 
\\Homi Bhabha Road, Bombay 400005, India. \\}
\date{\today}
\maketitle
\widetext
\begin{abstract}
We study the distribution of residence time or equivalently that of
``mean magnetization" for a family of Gaussian Markov processes indexed
by a positive parameter $\a$. The persistence exponent for 
these processes is simply given by $\theta=\a$ but the residence time
distribution is nontrivial. The
shape of this distribution undergoes a qualitative
change as $\theta$ increases, indicating a sharp change in the ergodic
properties of the process. We develop two alternate methods to
calculate exactly but recursively the moments of the distribution for
arbitrary $\a$. For some special values of $\a$, we obtain closed form
expressions of the distribution function.

\end{abstract}


\narrowtext

\section{Introduction}

The problem of {\it persistence} in spatially extended nonequilibrium
systems has recently generated a lot of
interest both
theoretically\cite{derrida1,derrida2,bray,majsire,diffusion,majbray,watson}
and experimentally\cite{marcos,yurke,tam}. These
systems include the Ising or Potts model undergoing
phase ordering 
dynamics\cite{derrida1,derrida2,bray,majsire,crit,klaus,clement,dg,krapiv,majcor},
simple diffusion equation with random
initial conditions\cite{diffusion,majbray}, several reaction diffusion
systems in both pure\cite{cardy} and disordered\cite{fisher} environments, 
fluctuating interfaces\cite{krug,kallabis,tnd},
Lotka-Volterra models\cite{frache} and inelastic collapse of
a randomly forced particle\cite{swift}. In many
of these systems the spatial degrees of freedom of the original many body
problem can be integrated out and the problem of persistence effectively
reduces to the calculation of the probability $P_0(t)$ of no zero crossing
upto some time $t$ of an effective single site stochastic process $y(t)$.

In most cases of interests, this probability decays as a power law for
large time, $P_0(t)\sim t^{-\theta}$, where the persistence exponent
$\theta$ is nontrivial. This nontriviality can be traced back to the
fact that once the spatial degrees of freedom are integrated out, the
effective single site process $y(t)$ becomes non-Markovian. For a
non-Markovian process, it is well known that the calculation of any
history dependent quantity such as persistence (no zero crossing
probability) is extremely hard\cite{slepian,reviews}.
As an example, for the diffusion equation with random initial condition,
the effective single site process $y(t)$ is a Gaussian non-Markovian
process characterized by its two-time correlator, $\langle
y(t_1)y(t_2)\rangle = [4t_1t_2/(t_1+t_2)^2]^{d/4}$ where $d$ is the
spatial dimension\cite{diffusion}. Even for this simple case, the
corresponding persistence exponent $\theta$ is nontrivial and is
known only numerically and approximately by analytical
methods\cite{diffusion,newman} but not exactly so far. Though recently,
an exact series expansion result for the exponent $\theta$ has been
derived for arbitrary smooth Gaussian 
\break{0.9in}
processes that includes the diffusion equation \cite{majbray}.

Recently it was argued\cite{dornic,newman} that given this stochastic
process $y(t)$, it might
be useful to investigate a more general quantity, namely the ``residence
time distribution", whose limiting behaviour determines the persistence
exponent. This is the distribution $f(r,t)$ of the random variable,
$r(t)={1\over
{t}}\int_0^t \theta [y(t')]dt'$ where $\theta (x)$ is the Heaviside
theta function. Thus $r(t)$ is simply the fraction of time spent by
the process $y(t)$ within time $t$ on one side of zero. It was
shown in ref. \cite{newman} that for any Gaussian stochastic process, the
distribution $f(r)$ is independent of time $t$. For Gaussian processes
with zero mean, the symmetry $r \leftrightarrow (1-r)$ indicates that
the function 
$f(r)$ is symmetric around $r=1/2$. Also in the limit $r\to 0$ (and 
symmetrically for $r\to 1$), the function $f(r)$ is clearly 
the probability that the process remains only on one side of zero and
hence is proportional to persistence. This indicates that as $r\to 0$,
the function $f(r)$ must behave as $\sim r^{\theta-1}$ (and as
$\sim (1-r)^{\theta-1}$ for $r\to 1$), so that $f(r)dr \sim t^{-\theta}$
as $r\to 0$ or $1$. A somewhat more convenient variable is the ``mean
magnetization"\cite{dornic},
$m(t)=2r(t)-1$, whose range is $[-1,1]$ and whose distribution
function, $P(m)={1\over {2}}f[(1+m)/2]$ is symmetric around $m=0$
and behaves as $P(m)\sim (1\pm m)^{\theta-1}$ near $m=\pm 1$.   

The distribution $P(m)$ is known exactly for the process that represents
the position of a one dimensional Brownian walker\cite{feller}.
Lamperti\cite{lamperti} derived an exact expression of $P(m)$ for
a class of renewal processes where successive zero crossing intervals are
statistically independent. Recently a special case of Lamperti's
results\cite{lamperti}, when the successive intervals are distributed
according to a Le\'vy law, was rederived by Baldassari et.
al.\cite{bald} by a different method.
The distribution $P(m)$ has been determined numerically for diffusion
equation\cite{newman} and
for interface growth models\cite{tnd}. Besides, moments of $P(m)$ have
been determined analytically for diffusion equation under the
independent interval approximation\cite{dornic}.

The distribution function $P(m)$ provides a somewhat more detailed
information on the statistical nature of the stochastic process $y(t)$.
For example, in the context of diffusion equation it was pointed out by
Newman and Toroczkai\cite{newman} that an interesting information can be
extracted from the shape of the function $P(m)$. For diffusion equation,
the exponent $\theta(d)$ (which controls the shape of the function $P(m)$
near $m=\pm 1$) increases monotonically with space dimension $d$.
There exists a critical dimension $d_c$ where $\theta(d_c)=1$ such
that for $d<d_c$, $\theta<1$ and the function $P(m)$ diverges as $m\to \pm
1$, has a minimum at $m=0$ and is concave upwards in the range $[-1,1]$.
On the other hand, for $d>d_c$, $\theta>1$, the function $P(m)$ goes
to zero as $m\to \pm 1$, has a maximum at $m=0$ and is convex upwards in
$[-1,1]$. The peak of the
distribution shifts from the edges $m=\pm 1$ to the center $m=0$ as $d$
increases through $d_c$. 
Thus for $d<d_c$, the most probable configurations of the
process $y(t)$ are the ones which do not cross zero whereas such
configurations are least probable for $d>d_c$, signalling the
existence of a sharp change in the ergodic properties of the diffusion
field. Such a detailed information is not contained in the
persistence exponent $\theta$. In ref. \cite{newman}, $d_c$ for diffusion
equation was approximately determined, $d_c\approx 36$.  

These useful informations contained in
$P(m)$ of the diffusion equation have so far not been possible to
derive exactly mainly
due to the non-Markovian nature of the single site Gaussian process.
It would therefore be useful to find and study some
simpler Markovian Gaussian processes with some tunable parameter
(which would play the
similar role as the spatial dimension $d$ does in diffusion equation)
where exact calculations can be performed. In this paper we study the
magnetization distribution $P(m)$ of a family of such Gaussian Markov   
processes parametrized by an index $\alpha$. By varying this
parameter $\alpha$, the persistence exponent $\theta$ for this process
can be varied continuously. The Markovian nature of the process also
makes many exact calculations possible.

The Markov process $y(t)$ that we study in this paper, satisfies the
following stochastic Langevin equation,
\bea
\frac{dy}{dt}=\sqrt{2 \a} t^{\a-1/2} \eta(t)  
\label{proc}
\eea
where $\eta(t)$ is a Gaussian white noise with $\langle \eta(t)\rangle =0$
and $\langle \eta(t_1)\eta(t_2)\rangle =\delta (t_1-t_2)$ and $\a$ is a 
positive parameter. There are various physical processes
that are described by the above Langevin equation. For example, for
$\a=1/2$, $y(t)$ represents the position of a one dimensional
Brownian random walker. For $\a=1/4$, $y(t)$ can be interpreted\cite{crit}
to be the ``total magnetization" of a Glauber chain undergoing
zero temperature coarsening dynamics after being quenched rapidly from
infinite temperature. 

The persistence of the process $y(t)$ is simply the probability for
this process not to cross zero upto time $t$ and decays as $t^{-\theta}$
for large $t$. The exponent $\theta$ for this process can be trivially
computed, $\theta=\a$. The simplest way to derive this is to define
a new time variable $t'=t^{2\a}$ such that the equation of motion
becomes $dy/dt' =\zeta (t')$ where the new noise $\zeta (t')$ has
zero mean and $\langle \zeta (t_1')\zeta (t_2')\rangle =\delta
(t_1'-t_2')$. But this is simply the equation of motion of a one
dimensional Brownian walker whose probability of no return to zero
upto time $t'$ decays as $\sim {1/{\sqrt {t'}}}\sim t^{-\a}$. Thus
the persistence exponent for $y(t)$ is simply $\theta=\a$. 

However we show in the rest of the paper that even though the
persistence exponent
$\theta=\a$ trivially for this process, the magnetization distribution
$P(m)$ is nontrivial. In fact, as $\alpha$ and hence $\theta$ is
increased, the shape of $P(m)$ changes from concave upwards to
convex upwards. For $\a=1/2$ (i.e., for ordinary
Brownian walker), the distribution $P(m)$ was already known
exactly, $P(m) = 1/({\pi \sqrt {1-m^2}})$\cite{feller}. For general $\a$,
while we have not been able to determine the full distribution
$P(m)$ in closed form, we demonstrate below by two completely different
methods that the moments of $P(m)$
can be calculated exactly. In the first method we generalize the formalism
developed by Kac\cite{kac} for $\a=1/2$ case to arbitrary $\a$. In the
second method, we use the formalism recently developed in the context of 
diffusion equation by Dornic and
Godr\`eche\cite{dornic} using independent interval
approximation (IIA). We point out however that while this latter method
yields only approximate results for diffusion equation\cite{dornic}, it
gives exact results for the Markov processes that we study in this paper. 
   
The paper is organized as follows. In section (II), we generalize
Kac's formalism for $\a=1/2$ to arbitrary $\a$ and derive an exact
recursion relations satisfied by the
moments of $P(m)$. In section (III), we rederive the
same results by using an alternate IIA formalism. In
section (IV) we use the formalism developed in sections-II and III to
obtain explicit results for the distribution of mean magnetization
for some special values of the parameter $\a$. Finally we conclude with
a summary and discuss the relative merits of
the two formalisms and some applications. 

\section{Method I: Generalization of Kac's formalism}

We consider the Gaussian process $y(t)$ evolving stochastically via
Eq. \ref{proc} and define the ``mean magnetization", $m(t)={1\over
{t}}\int_0^{t} V[y(t')]dt'$, where the functional $V(y)$ in our case
is simply, $V(y)=sgn(y)$. Let $G(y,t\mid y',t')$ denote the
propagator of the process, i.e., the probability that the process 
takes the value $y$ at time $t$ given that it was at $y'$ at time 
$t'<t$. This can be easily computed for our process and is given
by,
\bea
G(y,t\mid y',t')= {1\over {\sqrt {2\pi
(t^{2\a}-{t'}^{2\a})}}}e^{-{(y-y')^2/
{2(t^{2\a}-{t'}^{2\a})}}}.
\eea

Following Kac\cite{kac}, we define the moment generating function, 
\bea
\langle e^{-utm}\rangle =\sum_{n=0}^{\infty}
\frac{(-u)^n}{n!} \nu_n,
\label{nu0} 
\eea
where $\nu_n$ are the moments defined by:
\bea
\nu_n= \langle (\int_0^t V[y(t')]dt')^n \rangle.
\eea
To compute the moments $\nu_n$, it is useful to first define a set of
functions $Q_n(y,t)$ via the recursion relation,
\bea
Q_{n+1}(y,t) &=& \int_0^t dt'\int_{-\infty}^{\infty}dy'
G(y,t \mid y',t') V(y') Q_n (y',t')       \nonumber \\
Q_0(y,t) &=& G(y,t \mid 0,0).   \label{qdef}
\eea
It can then easily be checked that,
\bea
\nu_n=n!\int_{-\infty}^{\infty}Q_n(y,t)dy. 
\label{nu1}
\eea
Using Eq. \ref{nu0} and Eq. \ref{nu1}, we finally get,
\bea
\langle e^{-utm}\rangle = \int_{-\infty}^{\infty}dy Q_u(y,t)
\label{mom}
\eea
where $Q_u(y,t)$ is the generating function,
\bea
Q_u(y,t)=\sum_0^{\infty} Q_n (y,t) (-u)^n. 
\label{qudef}
\eea

Thus the moments of the mean magnetization $m$ can be computed exactly
from Eq. \ref{mom} provided we can evaluate the function $Q_u(y,t)$.
By using the recursion relation Eq. \ref{qdef}, it can be checked that
$Q_u(y,t)$ satisfies the following integral equation,
\bea
Q_u(y,t) = G(y,t\mid 0,0) ~~~~~~~~~~~~~~~~~~~~~~~~~~~~\nonumber \\
 -u\int_0^t dt' \int_{-\infty}^{\infty} dy' 
G(y, t \mid y', t') V(y') Q_u(y', t').
\eea
Using the definition of the propagator $G$, this integral equation can
then be converted to a partial differential equation,
\bea
\frac{\partial Q_u(y,t)}{\partial t}=\alpha t^{2 \alpha-1} \frac{\partial ^2
Q_u(y,t)}{\partial y^2}-u V(y) Q_u(y,t) 
\label{qproc}
\eea
with $Q_u(y,t=0)=\delta(y)$ and $V(y)=sgn(y)$. 

We first make the scale transforms:
\bea
z=\frac{y}{t^\alpha};~~~~~a=ut; ~~~~~Q_u(y,t)=\frac{1}{t^\a} F(z,a). 
\nonumber
\eea
Substituting in Eq. \ref{qproc} we get the following equation for $F$:
\bea
a\frac{\p F}{\p a} &=& \a \frac{\p ^2 F}{\p z^2}+\a z \frac{\p F}{\p z}
+[\a-a V(z)] F \nonumber \\
F(z,a=0) &=& \frac{e^{-z^2/2}}{\sqrt{2 \pi}}
\label{efeq}
\eea
where $V(z)=sgn(z)$. This equation has the following series solution:
\bea
F(z,a) &=& {1\over {\sqrt {2\pi}}}\sum_{n=0}^{\infty} b_n a^n e^a
D_{-n/\a}(-z)
e^{-z^2/4}~~~~~z<0 
\nonumber \\
F(z,a) &=& {1\over {\sqrt {2\pi}}}\sum_{n=0}^{\infty} c_n a^n e^{-a}
D_{-n/\a}(z)
e^{-z^2/4}~~~~~z>0, 
\label{sers}
\eea
where $D_p(z)$ are parabolic cylinder functions.
The coefficients $b_n$ and $c_n$ are to be determined from the
boundary conditions, namely the continuity of both $F$ and $\p F/{\p z}$ 
at $z=0$. The initial conditions determine $b_0=c_0=1$. 
Using the boundary conditions we get:
\bea
\sum_{n=0}^{\infty} b_n a^n e^a D_{-n/\a}(0) &=& \sum_{n=0}^{\infty} c_n
a^n e^{-a} D_{-n/\a}(0) \nonumber \\ 
\sum_{n=0}^{\infty} b_n a^n e^a D_{-n/\a+1}(0) &=& -\sum_{n=0}^{\infty} c_n
a^n e^{-a} D_{-n/\a+1}(0).  
\eea
By expanding in powers of $a$ and equating coefficients of all
powers of $a$ we finally obtain the following recursions for the
coefficients $b_n$ and $c_n$
\bea
c_{n} &=& \sum_{m=0}^{n-1} \frac{(-1)^m c_m}{(n-m)!}
\frac{D_{-m/\a}(0)}{D_{-n/\a}(0)}
~~~~~~~~~~~~~~~~~~ (n~ \rm{odd}) \nonumber \\
   &=& -\sum_{m=0}^{n-1} \frac{(-1)^m c_m}{(n-m)!}
\frac{D_{-m/\a+1}(0)}{D_{-n/\a+1}(0)}
c_{n-m}~~~~~ (n~\rm{ even}) \nonumber \\
b_n &=& (-1)^n c_n 
\label{para}
\eea
where we have used few identities satisfied by the parabolic cylinder
functions\cite{grads}.
Now from Eq. \ref{mom} it follows that the moments $\mu_k=\langle
m^k\rangle$ satisfy, 
\bea
\int_{-\infty}^{\infty} F(z,a) dz=\sum_{k=0}^{\infty}
\frac{(-a)^k}{k!}  \mu_k.
\eea
Finally, substituting the series solution for $F(z,a)$ (Eq. \ref{sers}) 
in the above equation we obtain:
\bea
\mu_n={n!}{\sqrt {2\over {\pi}}}\sum_{m=0}^n \frac{(-1)^m c_m}{(n-m)!}
D_{-m/\a-1}(0)
\label{mag}
\eea
for the even moments, while the odd ones vanish. The coefficents $c_m$ 
are determined from Eq. \ref{para}. This thus gives
an iterative scheme to generate all moments of the required distribution.  

\section{An alternate derivation of the moments}

There is an alternate scheme to calculate the moments of the distribution
$P(m)$. This scheme assumes statistical independence of the
successive zero crossing intervals of the process $y(t)$ and was first
used by Dornic and Godr\`eche in the context of diffusion
equation\cite{dornic}.
We however stress that while this assumption is only approximate for
non-Markov processes such as
diffusion equation\cite{dornic}, it is however exact for Markov processes
such as the one we study in this paper. An additional complication in
our case arises due to the fact that the average distance between
zero-crossings vanishes. This is a standard result which is true for any
Gaussian Markov process\cite{rice}. In
our calculations we introduce this average distance between two
consequtive zeros $\langle l\rangle $ as a
cut-off parameter and then take the limit $\langle l \rangle \to 0$
in the end.

Consider a particular realization of the process $y(t)$ ending at time
$t$. Let at time $t$ the process $y$ have a positive sign. Let $t_n$
denote the time instant at which the $n$th zero-crossing takes place. Then
the mean magnetization is given by:
\bea
m &=& \frac{1}{t}((t-t_n)-(t_{n}-t_{n-1})+...)=1-2 \xi,~~~\rm{ where}
\nonumber \\
\xi &=& \frac{t_n}{t}-\frac{t_{n-1}}{t}+\frac{t_{n-2}}{t}+... \nonumber
\eea
Similarly if $y(t) < 0$ then we get $x=2 \xi-1$. We note that at any $t$,
the sign of $y$ can be positive and negative with equal probability.
Hence if we can find the distribution of $\xi$, that of $m$ can be
computed easily.  
Now in the logarithmic time variable $T=\log(t)$, we can write $\xi$
in the form: 
\bea
\xi &=& e^{-(T-T_n)}-e^{-(T-T_{n-1})}+e^{-(T-T_{n-2})}+... \nonumber \\
    &=& e^{-\lambda} (1-e^{-l_n}+e^{-l_n-l_{n-1}}+...) \nonumber \\
&=& e^{-\lambda} X_n~~~~~~ \rm{where}  \\
X_n &=& (1-e^{-l_n}+e^{-l_n-l_{n-1}}+...),
\eea
$l_n=T_n-T_{n-1}$ and $\lambda$ is the time from the last
zero-crossing to time $t$. The variables $X_n$ satisfy Kesten
recursion relations,
\bea
X_n=1-e^{-l_n} X_{n-1}.
\eea
One then assumes that the successive zero crossing intervals are
statistically independent. In the long time limit the distribution of
$X$ is determined by the following set of equations:
\bea
X &=& \eta (1-2 \xi) + (1-\eta)(2 \xi-1)  \nonumber \\
\xi &=& e^{-\lambda} X    \nonumber \\
X &=& 1-e^{-l}X, \nonumber
\eea
where $\eta$ is an independent  random variable that can take values
$0$ and $1$ with equal probabilities.
Since one can compute the distributions of $l$ and $\lambda$, it is
then straightforward though tedious to compute all the moments of the
mean magnetization, $\mu_k=\langle m^k \rangle$
recursively\cite{dornic}. 

As noted above, the mean distance between zero crossings, $\langle l
\rangle $ vanishes and we introduce this as a cut-off parameter. We
now show that the $\mu_n$ are actually independent of $\D$.  

We first note that the Laplace transforms of the distributions of $l$
and $\lambda$, which we denote by $\hat f(s)$ and $\hat q(s)$
respectively, are given by \cite{dornic}:
\bea
\hat f(s) &=& \frac{1-\D g(s)}{1+\D g(s)}    \nonumber \\
\hat q(s) &=& \frac{2 g(s)}{s (1+ \D g(s))}, 
\label{pqlnth}
\eea 
where $g(s) = s (1-s \hat A(s))/2$, and $\hat A(s)$ is defined as follows.  
Consider the normalized process $Y= y(t)/{\sqrt {\langle
y^2(t)\rangle}}$. In the logarithmic time, $T=\log (t)$, 
this has a stationary autocorrelator, 
$C(T=\mid T_1-T_2\mid)=e^{-\a \mid T \mid}$. Now consider the
autocorrelation function $A(T)$ of the ``signed" process,
$A(T)=\langle sgn (Y(0))sgn (Y(T))\rangle$. 
The quantity ${\hat A(s)}$ is just the Laplace transforms of $A(T)$.
Using the fact that $y(T)$ is Gaussian with a correlator $C(T)$, the
function $A(T)$ can be easily computed, $A(T)=
(2/{\pi}){\sin}^{-1}[C(T)]$. In our case, $C(T)=e^{-\a \mid T \mid}$
which finally gives,
\bea
{\hat A}_s ={1\over s}- {1\over {s\pi}}B\big [ {{s+\a}\over {2\a}},
{1\over {2}}\big ]
\label{auto}
\eea
where $B[a,b]$ is the standard Beta function.

It is now convenient to define the moments $r_n=<X^n>/(1+\D g_n)$.
Then taking the $n$th power of the equation $X=1-e^{-l} X$ and using
the expression for $\hat f(s)$ given in Eq. \ref{pqlnth}, it can be
shown that $r_n$'s are 
recursively generated through the following sets of equations:
\bea
2 r_{2n+1} &=& \sum_{k=0}^{2n} {{2n+1} \choose k} (-1)^k r_k \nonumber \\
2 g_{2n}r_{2n} &=& -\sum_{k=1}^{2n-1} {{2n} \choose k} (-1)^k g_k r_k
\label{eqmn}
\eea
with $r_0=1$. Note that all $r_n$'s are {\it independent of $\D$}.
The moments of $\xi$ are then given by:
\bea
<\xi^n>=\hat q(n) <X^n>=\frac{2 g_n r_n}{n}
\label{eqxi}
\eea
and are also independent of the cut-off $\D$.
Finally the non-vanishing even moments of $m$ can be obtained through:
\bea
\mu_{2n}=\langle (2 \xi-1)^{2n} \rangle
\label{mueqn}
\eea
and clearly do not depend on $\D$.

The final expressions of the first few even moments are as follows (
see Eq. (B.4) in the appendix of ref. \cite{dornic}),
\bea
\mu_2 &=& {\hat A}_1 \nonumber \\
\mu_4 &=& 1- {{(1-3{\hat A}_1+4{\hat A}_2)(1-3{\hat A}_3)}\over
{1-2{\hat A}_2}} \nonumber \\
...
\label{appen}
\eea

We have checked that the moments $\mu_n$'s calculated recursively by this
method are identical to those obtained by the first method in section-II.

\section{Moments for some special values of $\a$}

In this section, we use the formalisms developed in the previous two
sections to derive some explicit results for the moments of the
distribution $P(m)$. 
While the iterative schemes developed in the previous sections are exact,
it seems that for general $\a$
it is quite hard to obtain an exact closed form expression 
of $\mu_n$ for aribitrary $n$. They have to be determined only
recursively. However the equations simplify for some special values of the
parameter $\a$, for which not just the moments but the full distribution
$P(m)$ can be obtained explicitly.

In order to see that the peak of the distribution shifts
from $m=\pm 1$ for small $\a$ to $m=0$ for large $\a$, it is natural to
examine 
the two extreme limits $\a=0$ and $\a=\infty$ for which fortunately we can
obtain exact form of the distribution. Consider first $\a=0$.
In this case it is somewhat easier to consider the second
method used in section (III). It can be easily seen then that all
the $g_n$'s vanish while the moments $r_n$'s remain finite.
Thus from Eq. \ref{eqxi} all moments of $\xi$ vanish.
Hence from Eq. \ref{mueqn} we get $\mu_n=1$ for all even
$n$. The same result can also be derived via the first method of
section (II) by taking carefully the $\a\to 0$ limit in Eqns.
\ref{para} and \ref{mag}. Immediate inspection then
determines, 
\bea
P(m)= {1\over {2}}\big [ \delta (m-1) +\delta (m+1) \big ]
\eea
for $\a=0$. Now consider the other extreme limit, $\a =\infty$. In this
case, one finds from Eq. \ref{para} that 
$c_m=1/{m!}$. Then the series in Eq. \ref{mag} just reduces to the 
expansion of $(1-1)^n$. Hence $\mu_n=0$ for all $n$. 
This indicates that for $\a=\infty$
\bea
P(m)=\delta (m).
\eea

Another case where exact form of $P(m)$ can be obtained is for $\a=1/2$.
In this case, using the known values of the parabolic
cylinder functions, it is easy to compute the first few terms of the
series $\{c_n, n=0,1,2,\cdots \}=\{ 1, 1, 2, 5, 14, 42, 132,
429,\cdots \}$ from Eq. \ref{para}. We then
make an ansatz, $c_n= (2n)!/[n!(n+1)!]$ and verify from Eq.
\ref{para} that it is indeed the solution for arbitrary $n$.
Substituting this
in Eq. \ref{mag}, we get
\bea
\mu_{2n} (\a=1/2)= {(2n)!\over { (n!)^2 2^{2n}}},
\eea
and the odd moments are identically zero. A little inspection then 
shows that these are the moments of the distribution function,
\bea
P(m) = {1\over {\pi \sqrt {1-m^2}}}
\eea
with $m$ varying in $[-1,1]$. We thus reproduce the well
known\cite{feller} magnetization distribution for the ordinary random walk
($\a=1/2$). 

Unfortunately we were unable to get a closed form expression of $\mu_n$ 
for other values of $\a$. For example, for $\a=1/4$, we get by
solving Eq. \ref{para} the first few terms of 
the sequence, $\{c_n, n=0,1,2, \cdots\}=\{1, 3, 72, 3663, 292824,
32227002, \cdots\}$. We found however that this is not listed 
in the catalogue of known integer sequences\cite{sloan} and we could not
guess any formula for this sequence. 

Thus as expected, the distributions of mean magnetization show a
qualitative change in shape as $\a$ changes. As we go from small
$\a$ to large $\a$, the peak of the distribution shifts from the edges
to the center. This can be understood physically since for
small $\a$ the noise becomes small as time increases and the probability
of zero crossing becomes negligible. On the other hand, for large $\a$,
the noise increases with time and the magnetization keeps changing sign
and thus the most probable value gets peaked at $m=0$.     

While obtaining an exact form of $P(m)$ is difficult for general $\a$,
there is no problem in obtaining the exact values of the moments of
$P(m)$ by using the recursion relations and the known values of
the parabolic cylinder functions. In Fig. \ref{moms}
we plot the moments for $\a=1/4$, $1/2$ and $3/4$. 

\section{Conclusion}

In this paper, we have studied the distribution of residence times or
equivalently that of mean magnetization of a family of Gaussian markov
processes parametrized by an index $\a$ which takes values continuously
from $0$ to $\infty$. We have shown that the shape of the
distribution $P(m)$ undergoes a qualitative change as $\a$ is increased
from $0$ to $\infty$. For small $\a$, 
$P(m)$ has peaks at the edges $m=\pm 1$ and has a minimum at $m=0$ whereas
for large $\a$, the peak of the distribution shifts to $m=0$ with minima at
the edges $m=\pm 1$. This change in the ergodicity properties
of a stochatic process as one changes a parameter was first noted in
ref. \cite{newman} in the context of diffusion equation. The advantage of
the process studied here, apart from representing various physical
situations, is that the Markov nature of the process makes it possible to
derive many exact analytical results.

In this paper we have developed two alternate formalisms to compute
the moments of the residence times or mean magnetization. While both
methods yield exact results for the moments, they do so only recursively.
A closed form expression for the moments and hence that of the full
distribution is possible only for some special values of the parameter
$\a$ that characterizes the process. But unfortunately this special set of
solvable values of $\a$ turn out to be the same for both these methods.
Thus so far as the problem studied in this paper is concerned, both
these methods are on equal footing. However there are other problems
where the former method that generalizes Kac's formalism seems to
have an advantage over the second method. We briefly mention below
one such application. 

The general problem of a random walker in a space with moving boundaries
has been well studied and has lot of applications\cite{redner}. It would
be interesting to study the residence time distribution in such problems.
For example, consider a random walker moving in one dimension and ask what
is the distribution of the fraction of times spent by the walker in
the region bounded by $+\infty$ and a point $O$ that moves
deterministically as $x_O(t)$ where $x_O(t)$ is some arbitrary function of
$t$. For the special case when $x_O(t)=c\sqrt {t}$ where $c$ is a
constant, this problem can be solved by using the techniques presented
in section-II of this paper. The calculations will be similar except that
the potential $V(z)=sgn (z)$ as used in Eq. \ref{efeq} should be replaced
by $V(z)=\theta (z-c)$. The corresponding equations can be solved as
before except that now the boundary conditions are to be applied at
$z=c$. We note however that the second method illustrated in section-III
does not seem to be easily generalizable to solve this problem.

We conclude with one last remark. The magnetization distribution
$P(m)$ is a useful quantity to study
for a generic stochastic process and contain in it many useful
informations regarding ergodicity etc. However as is obvious from the
efforts of this paper, exact analytical calculation of $P(m)$ seems quite
nontrivial even for the simple Gaussian Markov processes studied here.
Thus at present, the only hope to compute $P(m)$ for non-Markov
processes which are richer and more abundant in nature, seems to be
via numerical or approximate methods.

\section*{Acknowledgements}
We thank D. Dhar and M. Barma for useful discussions.

\vspace{2cm}

\vbox{
\epsfxsize=8.0cm
\epsfysize=8.0cm
\epsffile{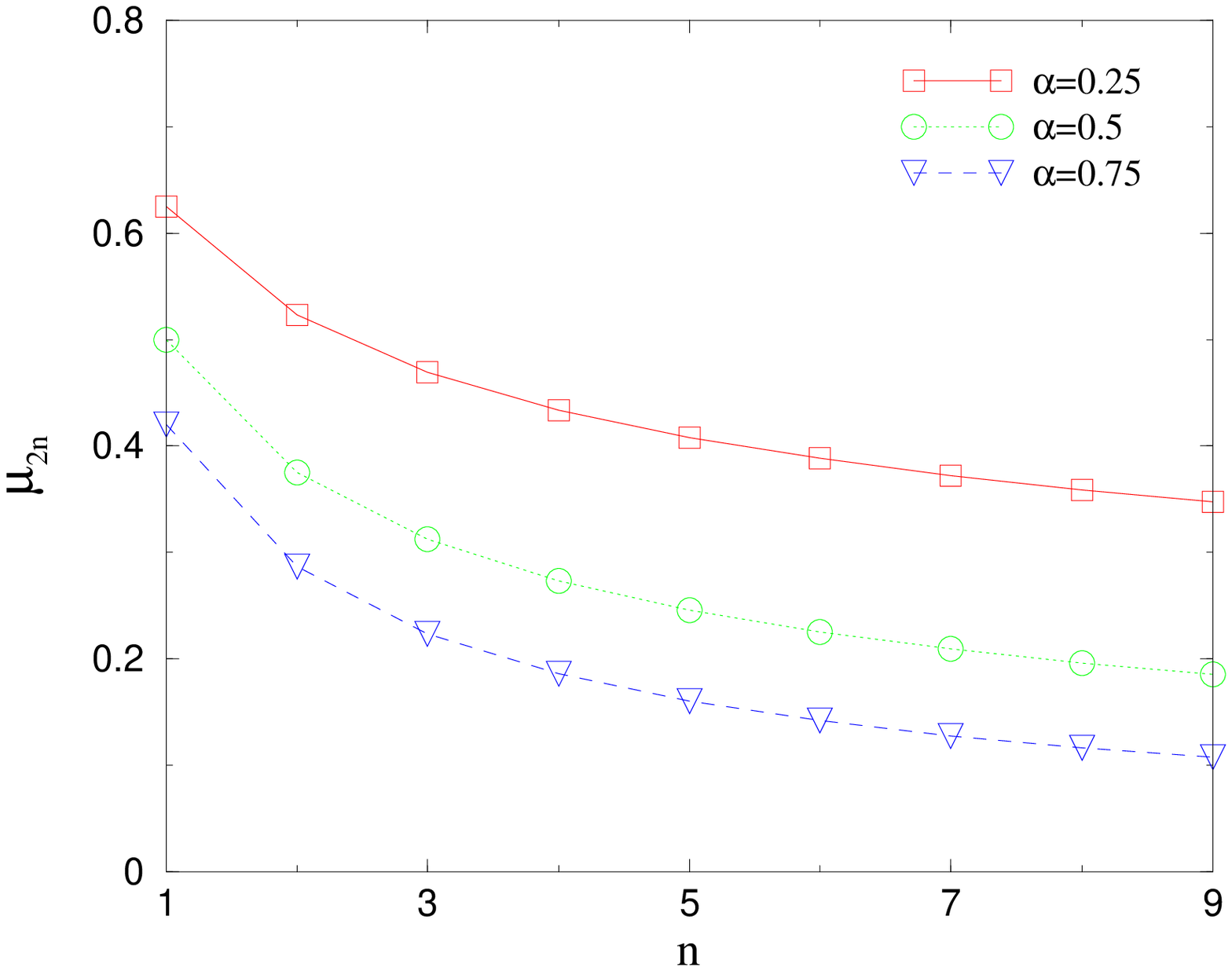}
\begin{figure}
\caption{\label{moms} In this figure the first few non-vanishing moments of 
$P(m)$ are plotted, for $\a=1/4$, $1/2$ and $3/4$. 
}  
\end{figure}}

\end{document}